\documentclass[aps,prl,twocolumn,superscriptaddress,showpacs]{revtex4-1}
\usepackage{graphicx}
\usepackage{color}
\usepackage{hyperref}

\hypersetup{colorlinks=false}

\begin{document}

\title{{Disorder-induced mutation of quasi-normal modes in 1D open systems}}

\author{Yury Bliokh}
\address{Department of Physics, Technion-Israel Institute of Technology, Haifa 32000,
Israel,}
\address{CEMS, RIKEN, Wako-shi, Saitama 351-0198, Japan}

\author{Valentin Freilikher}

\address{Department of Physics, Jack and Pearl Resnick Institute, Bar-Ilan
University, Israel}
\address{CEMS, RIKEN, Wako-shi, Saitama 351-0198, Japan}

\author{Franco Nori}
\address{CEMS, RIKEN, Wako-shi, Saitama 351-0198, Japan}
\address{Department of Physics, University of Michigan, Ann Arbor, Michigan
48109-1040, USA }

\begin{abstract}

We study the relation between quasi-normal modes (QNMs) and transmission
resonances (TRs) in one-dimensional (1D) disordered systems. We
show for the first time that while each maximum in the
transmission coefficient is always related to a QNM, the reverse statement
is not necessarily correct. There exists an intermediate state, where only
part of the QNMs are localized and these QNMs provide a resonant transmission. The rest
of the solutions of the eigenvalue problem (denoted as strange quasi-modes) are
never found in regular open cavities and resonators, and  arise exclusively due
to random scatterings. Although these strange QNMs belong to a discrete spectrum, they are
not localized and not associated with any anomalies in the transmission.
The ratio of the number of the normal QNMs to the total number of QNMs is independent of the type of disorder, and  deviates only slightly from the constant $\sqrt{2/5}$ in rather wide ranges of the strength of a single scattering and the length of the random sample.

\end{abstract}

\pacs{}

\maketitle

\address{Department of Physics, Technion-Israel Institute of Technology, Haifa 32000,
Israel,} \address{CEMS, RIKEN, Wako-shi, Saitama 351-0198, Japan}

\address{Department of Physics, Jack and Pearl Resnick Institute, Bar-Ilan
University, Israel} \address{CEMS, RIKEN, Wako-shi, Saitama 351-0198, Japan}

\address{CEMS, RIKEN, Wako-shi, Saitama 351-0198, Japan} 
\address{Department of Physics, University of Michigan, Ann Arbor, Michigan
48109-1040, USA }

\address{Department of Physics, Technion-Israel Institute of Technology, Haifa 32000,
Israel,} \address{CEMS, RIKEN, Wako-shi, Saitama 351-0198, Japan}

\address{Department of Physics, Jack and Pearl Resnick Institute, Bar-Ilan
University, Israel} \address{CEMS, RIKEN, Wako-shi, Saitama 351-0198, Japan}

\address{CEMS, RIKEN, Wako-shi, Saitama 351-0198, Japan} 
\address{Department of Physics, University of Michigan, Ann Arbor, Michigan
48109-1040, USA }



Wave processes in open systems can be described in terms of quasi-normal
modes (QNMs), which are a generalization of the notion of normal modes for
closed systems, to open structures, \cite{1,2,5,newJ,newL,newM,newR,newS,newT}. The corresponding
eigenfrequencies are complex, so that the imaginary parts characterize the
lifetime of the quasi-normal states. Regarding the transmission of  radiation through random media, it is more appropriate to use an alternative  approach  based on transmission resonances (TR):  open channels, through which the
radiation transmits with high efficiency \cite{3,5,we1,we2,6,7,8,8a,10,newA1,newF,newG,newN,newP}.

Recently, physicists came to realize that focusing radiation into such
channels could not only enhance the total intensity transmitted through
strongly-scattering media, but also: significantly improve images blurred by
random scattering, facilitate the detection and location of objects, provide
optical tomography at very large depths, etc.\cite{8,newF,newN,newG,newK}. To
efficiently excite  transmission resonances, it is preferable to treat them
as superpositions of QNMs, with which the incident signal can be coupled
by a properly-shaped wavefront \cite{8,8a}. The great potential of such
algorithms for a host of practical applications is obvious. 
 This is why the relation between  transmission resonances and QNMs have
recently attracted particular attention of both the physical \cite{11,12,13,7,Chen,Chaikin} and
mathematical communities \cite{1}.

It is now universally accepted that in open
systems (e.g., quantum potential wells, optical cavities, or microwave resonators)
each maximum in the transmission coefficient (i.e., transmission resonance) is
associated with a QNM, so that the resonant frequency is
close to the real part of the corresponding eigenvalue.  QNMs and TRs are often considered
identical. For example, the solutions of the eigenvalue problem (with no incoming
waves), which in physics are unambiguously called QNMs, in the mathematical
community dealing with the scattering inverse problem, are termed
transmission eigenvalues \cite{1}. However, the
connection between QNMs and TRs is not that simple and, despite extensive
research and much recent progress, still needs a better physical understanding
and  mathematical justification, at least for disordered systems.

To this end, it is instructive to look for insights the 1D limit
because its spectral and transport properties are better understood. It is
well-known \cite{Lifshitz} that the transmission of a long enough 1D disordered
system is typically (for most of the frequencies) exponentially small. At the same time, there exists a set of
frequencies where the transmission coefficient has local maxima
(resonances in transmission), some of them close to one \cite{weA}. Each resonance
is a transmission channel and is always associated with a QNM determined in
a standard way as a solution with outgoing boundary conditions. The reverse
statement, that each QNM manifests itself as a transmission resonance,
although never has been questioned, is usually taken as obvious and
self-evident, perhaps because  it is always the
case  in all regular (homogeneous or periodic)
quantum-mechanical and optical open  structures.

Here we show, both numerically and analytically, that in
1D \textit{disordered} systems there exist two types of QNMs:
ordinary QNMs, that provide resonance transmission peaks, and \textquotedblleft
strange\textquotedblright\ QNMs unrelated to any anomalies in the
transmission spectrum. These strange modes exist exclusively due to random
scatterings and arise already in the ballistic regime with weak disorder.
Although they belong to the discrete spectrum, their eigenfunctions are not
localized. The  imaginary parts of the strange QNMs
eigenfrequencies vary with increasing  disorder in a highly
unusual manner. Indeed, typically, the stronger  the disorder is, the more
confined the system becomes, which
implies that the eigenfrequencies should approach the real axis. However, the
imaginary part of a strange mode's eigenfrequency either increases from the onset of
disorder, or goes down anomalously slowly. Most surprisingly, up to rather
strong disorder, the average ratio of the density (in the frequency domain)
of strange modes to the total density of QNMs, being independent of the type of disorder, remains close to the constant $\sqrt{2/5}$ in wide ranges of the strength of disorder and of the total length of the system. 
The value $\sqrt{2/5}$ follows from the
general statistical properties of random trigonometric polynomials \cite%
{Edelman}. As the disorder keeps growing, eventually all strange quasimodes
turn normal. Therefore these results can be interpreted as a manifestation
of the existence (in 1D random systems, at least) of an intermediate regime, at which in
any finite-frequency interval, only a part of the quasimodes are localized
and provide resonant transmission.


We consider a generic 1D system composed of $N+1$ scatterers separated by $N$
intervals and attached to two semi-infinite leads. Two problems are
associated with such systems. The first one is  finding
solutions $\psi (x,t)$ of the wave equation  satisfying  the outgoing boundary
conditions, which means that there are no right/left-propagating waves in
the left/right lead. The eigenfunction
solution $\psi_n(x,t)$ of this problem  is the superposition of two counter-propagating monochromatic waves $\psi_n(x)^{(\pm)}e^{-i\omega_n t}$. In any $j$th layer $\psi_n^{(\pm)}(x)=\psi_{n,j}^{(\pm)}(x)=a_{n,j}^{(\pm)}e^{\pm ik_nx}$. The amplitudes $a_{n,j}^{(\pm)}$ in adjacent layers are connected by a transfer matrix. 
The wave numbers $k_{n}$ are
complex-valued and form the discrete set (poles of the scattering matrix) $%
k_{n}^{(\mathrm{mod)}}=k_{n}^{\prime }-ik_{n}^{\prime \prime }$, $k^{\prime
\prime }>0,$ and frequencies $\omega _{n}^{(\mathrm{mod)}}=ck_{n}.$ The
corresponding eigenfunctions are the so-called QNMs.
Note that all distances hereafter are measured in optical lengths.
The second problem is the transmission of an incident wave through the
system. The set of  wave numbers and corresponding fields inside the
system for which the transmission coefficient reaches its local maximum are
the so-called TRs. Evidently these two problems are interrelated. In particular, the density of QNSs at a frequency $\omega$ is proportional to the derivative with respect to frequency of the phase of the complex transmission coefficient \cite{Avishai, Akkermans}.

The goal of this paper is to establish the relation between  the spectra and
wave functions of QNMs and TRs.

In what follows, the scatterers and the distances between them are
characterized by the reflection coefficients $r_{i}\equiv r_{0}+\delta r_{i}$
and lengths $d_{i}\equiv d_{0}+\delta d_{i}$, respectively. The random
values $\delta r_{i}$ and $\delta d_{i}$ are distributed in certain
intervals, and $\left\langle \delta r_{i}\right\rangle =0$ and $\left\langle
\delta d_{i}\right\rangle =0$. Here, $\left\langle\ldots\right\rangle$ stands
for the value averaged over the sample. The last condition means that the
total length $L$ of the system is equal to $Nd_{0}$ and therefore any random
realization with the same $N$ contains the same number of QNMs.

To explicitly introduce the tunable strength $s$ of disorder, we replace
all reflection coefficients, except for those at the left, $r_{L}$, and right, $%
r_{R}$, edges of the system by $sr_{i} $, and assume (unless otherwise 
specified) that the coefficients $r_{i}$ are homogeneously distributed in
the interval $(-1,1).$ This notation enables keeping track of the evolution
of the QNM eigenvalues $k_{n}^{(\mathrm{mod)}}$ and of the resonant wave
vectors $k^{(\mathrm{res})}$ when the disorder increases from zero ($s=0$) 
while the reflection
coefficients $r_{L}$ and $r_{R}$ at the semitransparent boundaries remain constant.

When $s=0$, (i.e., no disorder) the real and imaginary parts of the QNM eigenvalues $k_{n}^{(%
\mathrm{mod)}}$ are 
\begin{eqnarray}
k_{n}^{\prime } &=&\frac{1}{2L}\cdot \left\{ 
\begin{array}{r}
\pi +2\pi n,\hspace{5mm}\mathrm{when}\hspace{2mm}r_{L}r_{R}>0, \\ 
2\pi n,\hspace{5mm}\mathrm{when}\hspace{2mm}r_{L}r_{R}<0,%
\end{array}%
\right.   \label{eq1} \\
k_{n}^{\prime \prime } &=&-\frac{1}{2L}\ln |r_{L}r_{R}|.  \label{eq2}
\end{eqnarray}
The wave intensity, defined as $I_{n,j}=|\psi _{n,j}^{(+)}|^{2}+|\psi
_{n,j}^{(-)}|^{2}$ is distributed along the system as $%
I_{n}(x_{j})\propto \cosh [2k^{\prime \prime }(x_{j}-x^{\ast })]$, where $%
x^{\ast }=L[1-\ln (|r_{R}/r_{L}|)/\ln(|r_{R}r_{L}|)]/2$. When $|r_{L}|=|r_{R}|
$, the minimum of the intensity is located at the centre of the system, and
in an asymmetric case shifts to the boundary with a higher reflection
coefficient. This property will be used when analyzing the behavior of the
QNMs when the disorder parameter $s$ grows.

It is easy to show that when $s=0$ the wave numbers $k_{n}^{(\mathrm{res})}$
of the transmission resonances coincide with the real parts $k_{n}^{\prime }$
given by Eq.~(\ref{eq1}). Thus, in the homogeneous resonator, there is a
one-to-one correspondence between QNMs and TRs. The same correlation exists
also in periodic systems (periodic sets $r_{i}$ and $d_{i}$) \cite{Settimi}.

The question now is whether this relationship survives in the
disordered system, when $s\neq 0$. There is strong evidence \cite{5,6,7,8,10} that for every resonance there is a corresponding
QNM. However,  as we show below, the reverse statement is not valid: there
are certain QNMs which cannot be associated with any resonance.

\begin{figure}[tbh]
\centering \scalebox{0.325}{\includegraphics{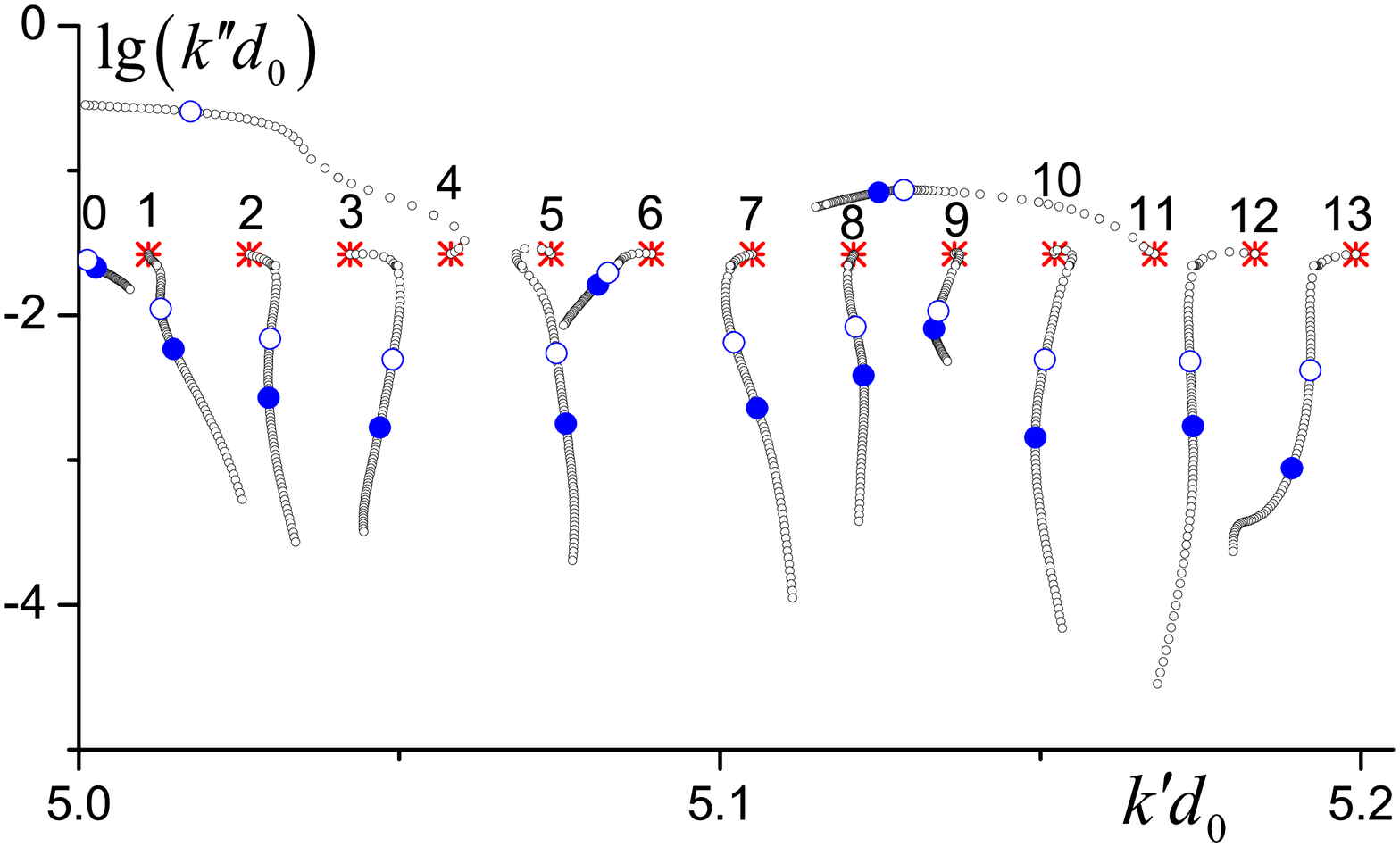}}
\caption{(color online) Motion of the QNMs eigenvalues, $k_{n}^{(\mathrm{mod)}%
}=k_{n}^{\prime }-ik_{n}^{\prime \prime }$, as the degree $s$ of disorder
grows. Red asterisks mark the initial positions with no disorder ($s=0$). Red open circles and blue solid circles show the positions of the QNMs eigenvalues 
at $s=0.1$ and $s=0.2$, respectively. Note that ordinary QNMs shift down when increasing disorder while some strange QNMs (e.g. \#4 and 11) shift up.}
\label{Fig1}
\end{figure}

Figure~\ref{Fig1} shows the evolution of the eigennumbers $k_{n}^{\mathrm{mod}%
} $ in the complex plane $(k^{\prime },k^{\prime \prime })$ as $s$ grows.
Initially, when $s=0$, all eigenumbers are equidistantly located on the line 
$k^{\prime \prime }={\rm const}$, in agreement with Eqs.~(\ref{eq1}, \ref{eq2}). As soon as disorder arises $(s\neq 0)$ and increases, the eigenvalues
separate into two essentially different types. Indeed, with $s$
increasing, the points \#1-3,5,7,8,10,12,13 in Fig. \ref{Fig1} move towards
the real axis ($k_{i}^{\prime \prime }$ decrease) with approximately the
same \textquotedblleft velocity\textquotedblright\ (ordinary QNMs). The rest
of the points (strange QNMs)  either shift down substantially more slowly (\#0,6,9)
or, even more surprisingly, move away from the real axis (points 4 and 11).
The latest modes are highly unusual because disorder makes them
more leaky. This is quite the opposite to the hitherto
observed and well understood increase of the lifetime of the eigenstates due
to multiple scattering. 

\begin{figure}[t]
\centering \scalebox{0.3}{\includegraphics{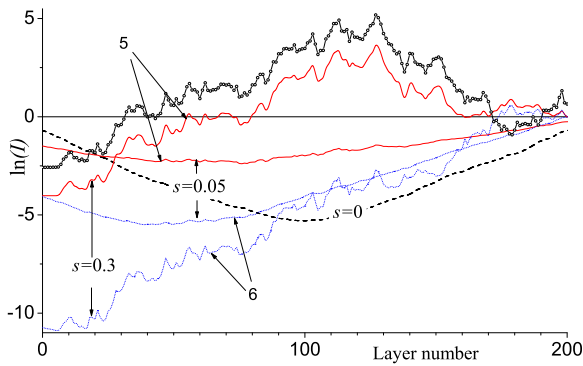}}
\caption{(color online) Spatial distribution of the intensity $I(j)$ of  QNMs \#5 (solid curves) and \#6
(dotted curves) for different values of the disorder strength $s$. The dashed black curve corresponds
to a homogeneous resonator ($s=0$, $r_{L}=-r_{R}=0.005$). The resonance intensity
distribution for $s=0.3$ is shown by circles. }
\label{Fig2}
\end{figure}

The difference between the ordinary and strange QNMs goes beyond the
evolution of the eigenvalues and manifests itself also in the the spatial
distribution of the QNM intensity inside the system. As an example, the
spatial distributions along the system of the intensities $I_{5,j}$ and $I_{6,j}$ of
QNMs \#5 and 6 are presented in Fig~\ref{Fig2}, for different
values of $s$. Note that the difference between the imaginary parts $k^{\prime
\prime }$ of the eigenvectors 5 and 6 increases as $s$ increases (see Fig.~\ref{Fig1}). 
Despite the fact that the initial ($s=0$) distributions are identical, even
small disorder ($s=0.05$) deforms the distributions $I_{5,j}$ and $I_{6,j}$
in very different ways. The distribution $I_{5,j}|_{s=0.05}$ is
similar to $I_{5,j}|_{s=0}$, but has a much less pronounced minimum. By
analogy with a homogeneous resonator, this can be interpreted as the growth of the
effective reflection coefficients $r_{L}$ and $r_{R}$, which agrees well
with the statement that the wave lifetime increases when disorder becomes
stronger. For larger $s$, $I_{5,j}$ tends to manifest the behaviour
typical for QNM in the localized regime. In contrast, the intensity
evolution of QNM \#6 is similar to that in the homogeneous resonator, whose
right wall becomes more transparent. QNMs \#4 and 11 also demonstrate the same
behaviour, but the effective transparency of one of the \textquotedblleft
walls\textquotedblright\ increases much faster when the degree of disorder $s$
grows.


We also consider the propagation of a monochromatic wave through the same
system. When $s=0$, the number of resonances $N_{\mathrm{res}}$ is equal to
the number of QNMs, $N_{\mathrm{mod}},$ and all $k_{n}^{(\mathrm{res})}$
coincide with the real parts $k_{n}^{\prime }$\ of QNMs. When disorder is
introduced, $s\neq 0$, each $k_{n}^{(\mathrm{res})}$ remains close to the $k_{n}^{\prime }$ of the corresponding ordinary QNM: $k_{n}^{(\mathrm{res}%
)}(s)\simeq k_{n}^{\prime }(s)$. The spatial intensity distributions of QNM
\#5 and of the corresponding TR are also similar, up to small details (see Fig.~%
\ref{Fig2}).

\begin{figure}[tbh]
\centering \scalebox{0.32}{\includegraphics{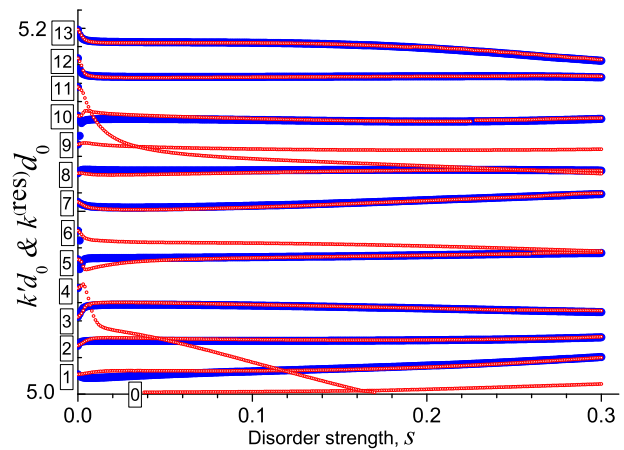}}
\caption{(color online) Dependencies $k_{\mathrm{res}}(s)$ (thick blue solid circles) and $%
k^{\prime }(s)$ (thin open red circles). QNMs are numbered as in Fig.~\ref{Fig2}. It is seen that for ordinary QNMs, $k^{\mathrm{%
res}}(s)$ and $k^{\prime }(s)$ practically coincide, whereas there are no
resonances associated with strange QNMs (\#0,4,6,9,11).}
\label{Fig3}
\end{figure}

However, the transmission resonances whose frequencies at $s=0$ are
equal to the real parts of the eigenvalues of the strange QNMs, disappear
when the mean value of the reflection coefficients $s\left\langle
r_{i}\right\rangle $ becomes of the order of $r_{L,R} $. Figure~\ref{Fig3}
demonstrates this behavior. Here, the role of the reflection coefficients $%
r_{R,L}$ ($r_{R}=r_{L}=0.005$ in Figs.~\ref{Fig2} and \ref{Fig3}) is only to specify the TRs
and QNMs at $s=0.$ 

\begin{figure}[tbh]
\centering \scalebox{0.32}{\includegraphics{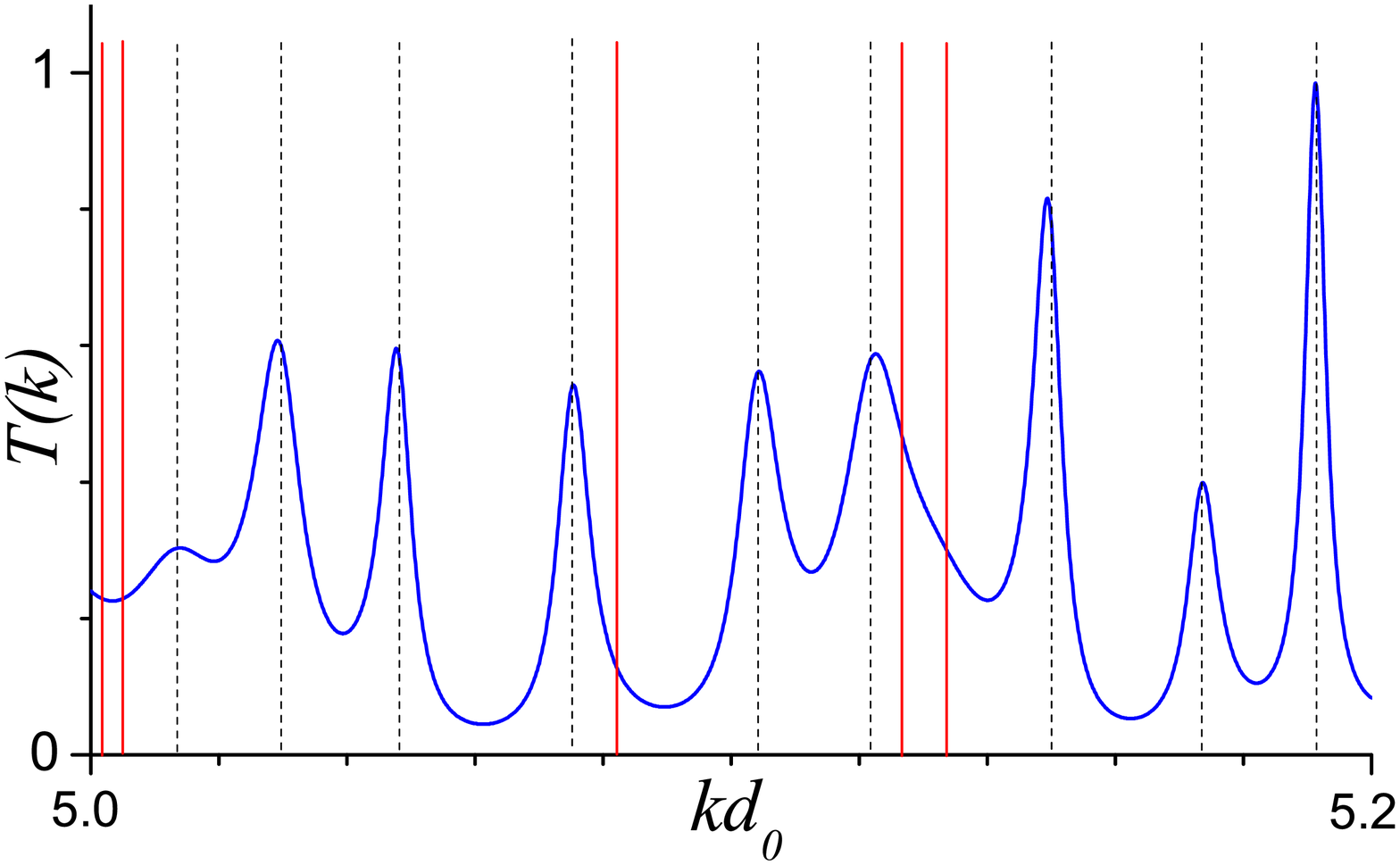}}
\caption{(color online) Transmission spectrum $T(k)$ at $s=0.15$. The black dashed (red solid) vertical lines
indicate the $k^\prime_n$ values of ordinary (strange) QNMs.}
\label{FigA}
\end{figure}

Thus, any TR has its partner among QNMs, but the reverse is not true:
there are strange QNMs that are not associated with any maxima in the
transmission, as it is shown in Fig.~\ref{FigA}, and
therefore do not have co-partners between resonances. 
In other words, in a given wave number interval $\Delta k$, the statistically-averaged number of TRs, $N_{\mathrm{res}}$, is smaller than the statistically-averaged number of QNMs,  $N_{\mathrm{mod}}=\Delta kL/\pi $, and does not depend on the degree of disorder. This fact was noticed in the numerical calculations in \cite{Chaikin}.


Surprisingly, when $s\rightarrow 0$, the ratio $N_{\mathrm{res}}/N_{\mathrm{mod}}$ is a \textit{%
universal} constant $\sqrt{2/5}$, independent of the type of disorder,  and remains practically
independent on the degree of disorder and  the length $L$ of the
system in a rather broad range of these parameters.

\begin{figure}[tbh]
\centering \scalebox{0.32}{\includegraphics{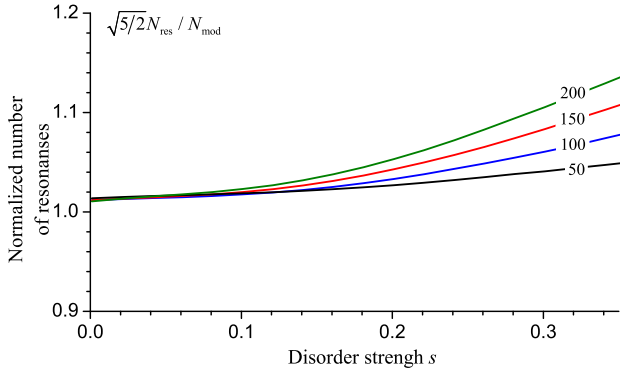}}
\caption{(color online) Normalized ratio $\protect\sqrt{5/2}N_{\mathrm{res}}/N_{%
\mathrm{mod}}$ versus the degree of disorder  $s$ for systems of various lengths 
$L=Nd_{0}$ ($N$ is the number of layers). }
\label{Fig4}
\end{figure}

Figure \ref{Fig4} shows the ratio $N_{\mathrm{res}}/N_{\mathrm{mod}}$ as a function of $s$,
statistically averaged over $10^{4}$ random realizations and normalized by
$\sqrt{2/5}$,  for various lengths $L$ in
the case $r_{L,R}=0$. 
It is important to note that the localization length (measured in 
numbers of layers) $N_{\mathrm{loc}}\propto s^{-2}$, and this is less than 20 for $%
s=0.3$. This means that $N_{\mathrm{res}}/N_{\mathrm{mod}}\simeq \sqrt{2/5}$
even when the system dimension exceeds considerably the localization length.


Figure~\ref{Fig3} shows that the difference betweens $N_{\mathrm{res}}$ and $%
N_{\mathrm{mod}}$ appears when $s$ is very small so that $N_{\mathrm{loc}%
}\gg N$, and remains practically unchanged even when $s$ is rather large so
that $N_{\mathrm{loc}}\ll N$. This means that the origin of this phenomenon is
not specifically related to  localization and can be studied when $s$ is
arbitrarily small.

To calculate the average number of  TRs in the limit $s\ll 1$, we use the single-scattering approximation and write the
total reflection coefficient $r(k)$ of the whole system as: 
\begin{equation}
r(k)=\Sigma _{n=1}^{N}r_{n}e^{2ikx_{n}},  \label{eq3}
\end{equation}%
where $x_{n}$ is the coordinate of the $n$-th scatterer. The values $k_{\max
}$, where the transmission coefficient, $T(k)=1-|r(k)|^{2}$, has local
maxima, are defined as the zeros of the function $f(k)\equiv d|r(k)|^{2}/dk=2%
\mathrm{Re}\left[ r(k)dr^{\ast }(k)/dk\right] $: 
\begin{equation}
f(k_{\max })=4\mathrm{Im}\Sigma _{n=1}^{N}\Sigma _{m=1}^{N}r_{n}r_{m}x_{m}e^{2ik_{\max
}\left( x_{n}-x_{m}\right)}=0.
\end{equation}

Assuming first  that $\delta d_{i}=0$, then $f(k)$ becomes 
\begin{eqnarray}\label{eq5}
f(k)\propto\Sigma _{n=1}^{N}\Sigma _{m=1}^{N}r_{n}r_{m}(m-n)\sin \left[
2k(m-n)d_{0}\right]\nonumber\\
=\Sigma _{l=1}^{N}\sin \left( 2kld_{0}\right) \left\{ \Sigma
_{n=1}^{N-l}r_{n+l}r_{n}l\right.\nonumber \\
\left. +\Sigma _{n=l}^{N}r_{n-l}r_{n}l\right\}\equiv\Sigma
_{l=1}^{N}\sin \left( 2kld_{0}\right) a_{l}.
\end{eqnarray}%
Eq.~(\ref{eq5}) is the trigonometric sum $\Sigma
_{l=1}^{N}a_{l}\sin \left( \nu _{l}k\right) $ with \textquotedblleft
frequencies\textquotedblright\ $\nu _{l}=2ld_{0}$ and random coefficients $%
a_{l}$. The statistics of the zeroes of random polynomials have been studied
in \cite{Edelman}, where it is shown that the statistically-averaged number of
real roots $N_{\mathrm{root}}$ of the sum of this type at a certain interval 
$\Delta k$ is 
\begin{equation}
N_{\mathrm{root}}=\frac{\Delta k}{\pi }\sqrt{\frac{\Sigma \nu
_{l}^{2}\sigma _{l}^{2}}{\Sigma \sigma _{l}^{2}}},  \label{eq7}
\end{equation}%
where $\sigma _{l}^{2}=\mathrm{Var}(a_{l})$ is the variance of the
coefficients $a_{L}=\Sigma _{n=1}^{N-l}r_{n+l}r_{n}l+\Sigma
_{n=l}^{N}r_{n-l}r_{n}l$. When $N\gg 1$, 
\begin{equation}
\mathrm{Var}(a_{l})\simeq 2(N-l)l^{2}\sigma _{0}^{4},  \label{eq9}
\end{equation}%
where $\sigma _{0}^{2}=\mathrm{Var}(r)$. The sums in Eq.~(\ref{eq7}) can be
calculated using Eq.~(\ref{eq9}), which yields \cite{GR}: 
\begin{eqnarray}
\Sigma _{l=1}^{N}\sigma _{l}^{2} &=&2\sigma _{0}^{4}\Sigma
_{l=1}^{N}l^{2}(N-l)\simeq \frac{1}{6}\sigma _{0}^{4}N^{4},  \nonumber \\
\Sigma _{l=1}^{N}\nu _{l}^{2}\sigma _{l}^{2} &=&8d_{0}^{2}\Sigma
_{l=1}^{N}\sigma _{0}^{4}l^{4}(N-l)\simeq \frac{4}{15}d_{0}^{2}N^{6}\sigma
_{0}^{4}.  \label{eq10}
\end{eqnarray}%
From Eqs.~(\ref{eq7}) and (\ref{eq10}) we obtain{\Large \ } 
\begin{equation}
N_{\mathrm{root}}=\frac{2\Delta kNd_{0}}{\pi }\sqrt{\frac{2}{5}}=2\frac{%
\Delta kL}{\pi }\sqrt{\frac{2}{5}},  \label{eq12}
\end{equation}%
where $L=Nd_{0}$. Since the number of minima of
the reflection coefficient is equal to the number of TRs, $N_{\mathrm{res}%
}=N_{\mathrm{root}}/2$, and the number $N_{\mathrm{mod}}$ of QNMs in the
same interval $\Delta k$ is $N_{\mathrm{res}}=\Delta kL/\pi $, from Eq. \ref%
{eq12} it follows that 
\begin{equation}
{N_{\mathrm{res}}}/{N_{\mathrm{mod}}}=\sqrt{{2}/{5}}  \label{eq13}
\end{equation}%

This analytically-calculated relation agrees perfectly with the results of
numerical calculations performed \textit{without} assuming any periodicity of the scatterers. 
To calculate this ratio for more general situations, when the distances
between the scatterers are also random ($\delta d_{i}\neq 0),$ the
frequencies $\nu =2ld_{d}$ in Eq.~(\ref{eq5}) should be replaced by $\nu
=2|x_{m}-x_{m\pm l}|$. Since the main contribution to the sums in Eq.~(\ref%
{eq7}) is given by the terms with large $l\sim N$, the mean value of $%
|x_{m}-x_{m\pm l}|$ can be replaced by $ld_{0}$, in the case of a homogeneous
distribution of the distances $d_{n}$ along the system. This
ultimately leads to the same result Eq. (\ref{eq13}).


In summary, it is well known that there is a one-to-one correspondence between the
QNMs of a regular open system (wave resonator or quantum cavity) and
its transmission resonances: each QNM is unambiguously associated with a TR,
and vice versa. In this paper, we show for the first time that in 1D random
structures, this reciprocity is
broken: any weak disorder mutates  part of the eigenstates so that
the corresponding resonances in the transmission disappear and the density of TR
becomes smaller than the total density of states. Although the strange
modes belong to a discrete spectrum, the spatial structure of the
eigenfunctions differs drastically from that of the ordinary states and
show no sign of localization. It is significant that while  the strange modes do not show up in the amplitude of the transmission coefficient, in the phase of the transmitted field they manifest themselves in just the same way as the ordinary modes do. Indeed, the numerical calculations show that the derivative of this phase with respect to the frequency gives the total density of QNMs, which includes both the normal ordinary and strange ones. When the disorder is weak (but strong enough
to localize the ordinary modes), the ratio of the number of TRs to the
total number of QNMs in a frequency interval $\Delta \omega \rightarrow
\infty $  is independent of the type of disorder and anomalously weakly deviates from a universal constant, $\sqrt{2/5}$, when the strength of disorder and the length of the random sample increase. This constant coincides with the one
analytically calculated in the weak single-scattering approximation
ensemble-averaged ratio $N_{\mathrm{res}}/N_{\mathrm{mod}}$. If the
strength $s$ of disorder grows, ultimately all strange quasimodes become
ordinary. This means that in 1D random systems there exists an intermediate,
so far unknown regime, at which in any finite-frequency interval, only a
part of the quasimodes are localized and provide resonant transmission.

\begin{acknowledgments}

We gratefully acknowledge stimulating discussion with K. Bliokh. We specially thank M. Dennis who drew our attention to the paper \cite{Edelman}.

This research is partially supported by the
RIKEN iTHES Project,
MURI Center for Dynamic Magneto-Optics,
and a Grant-in-Aid for Scientific Research (S).

\end{acknowledgments}


\begin{thebibliography}{99}
\bibitem{1} F. Cakoni, H. Haddar (Editors), Inverse Problems \textbf{29},
Topical Issue, (2013).

\bibitem{2} E.S.C. Ching, P.T. Leung, A. Maassen van den Brink, W.M. Suen, S.S. Tong, and K. Young, Rev. Mod. Phys. \textbf{70}, 1545 (1998).




\bibitem{5} J. Wang and A. Genack, Nature \textbf{471}, 345 (2011).

\bibitem{newJ} N.Hatano and G. Ordonez, arXiv:1405.6683 (2014).

\bibitem{newL}  Wonjun Choi, Q-Han Park, and  Wonshik Choi, Opt. Exp. \textbf{20}, 20721 (2012).

\bibitem{newM} C. Sauvan, J.P. Hugonin, I.S. Maksymov, and P. Lalanne, arXiv:1304.8110.

\bibitem{newR} C. Vanneste and P. Sebbah, Phys. Rev. A \textbf{79}, 041802
(2009).

\bibitem{newS} F.A. Pinheiro, M. Rusek, A. Orlowski, and B.  van Tiggelen,
Phys. Rev. E \textbf{69}, 026605 (2004). 

\bibitem{newT} P. Sebbah, B. Hu, J. Klosner, and A. Genack, Phys. Rev. Lett. \textbf{96},
183902 (2006).


\bibitem{3} O. Dorokhov, Solid State Com. \textbf{44}, 915 (1982); \textit{ibid.} \textbf{51}, 381, (1984).


\bibitem{6} Wonjun Choi, A. Mosk, Q-Han Park, and Wonshik Choi, Phys. Rev. B 
\textbf{83}, 134207 (2011).

\bibitem{7} A. Pena, A. Girschik, F. Libisch, S. Rotter, and A. Chabanov,
Nature Communications \textbf{5}, 1 (2014).

\bibitem{8} S. Liew, S. Popoff, H. Cao, A. Mosk, and W. Vos, arXiv:1401.5805 (2014).

\bibitem{8a} {S.M. Popoff, A. Goetschy, S.F. Liew, A.D. Stone, and H. Cao, Phys. Rev. Lett. \textbf{112}, 133903 (2014).} 

\bibitem{10} X. Cheng, C. Tian, and A. Genack, Phys. Rev. B \textbf{88}, 094202 (2013).

\bibitem{newA1}  Z. Shi and A.Z. Genack, Phys. Rev. Lett. \textbf{108}, 043901 (2012). 

\bibitem{newG} I. Vellekoop and A. Mosk, Phys. Rev. Lett. \textbf{101}, 120601 (2008).

\bibitem{newN} Z. Shi, M. Davy, J. Wang, and A. Genack,  Opt. Lett. \textbf{38}, 2714 (2013).


\bibitem{newP} B. G\'{e}rardin, J. Laurent, A. Derode, C. Prada, and A. Aubry, 
arXiv:1404.2092 (2014).

\bibitem{newF} A. Mosk, A. Lagendijk, G. Lerosey, and M. Fink, Nature Phot. \textbf{6}, 283 (2012).

\bibitem{we1}  K. Bliokh, Y. Bliokh, V. Freilikher, S. Savel'ev, and Franko Nori, Rev. Mod. Phys. \textbf{80}, 1201 (2008).

\bibitem{we2} K. Bliokh, Y. Bliokh, V. Freilikher, and Franko Nori, ``Anderson Localization of Light in Layered Dielectric Structures'', in \textit{Optical properties of photonic structures: interplay of order and disorder}, ed. by. M. Limonov and R. De La Rue, (CRC Press) p.57 - 86 (2012).



\bibitem{newK} A. Goetschy and A. Stone, Phys. Rev. Lett. \textbf{111}, 063901 (2013).

\bibitem{11} W. L. Vos, T. W. Tukker, A. P. Mosk, A. Lagendijk, and W. L.
IJzerman, Appl. Opt. \textbf{52}, 2602 (2013).

\bibitem{12} M. Kim, W. Choi, C. Yoon, G.H. Kim, and W. Choi, Opt. Lett. 
\textbf{38}, 2994 (2013).

\bibitem{13} Wonjun Choi, M. Kim, D. Kim, C. Yoon, C. Fang-Yen, Q. Park, and
Wonshik Choi, arXiv:1308.6558 (2013).

\bibitem{Chen} {L. Chen, C. Lv, and X. Jiang, Comp. Phys. Commun. \textbf{183}, 2513 (2012).}

\bibitem{Chaikin} Y. Bliokh, E. Chaikina, N. Liz\'{a}rraga, E. Mendez, V. Freilikher, and Franco Nori, Phys. Rev B \textbf{86}, 054204 (2012).

\bibitem{Lifshitz}  I. Lifshitz, S. Gredeskul, L. Pastur,\textit{
Introduction to the Theory of Disordered Systems} (Wiley, New York, 1989).

\bibitem{weA} K. Yu. Bliokh, Yu. P. Bliokh, and V. D. Freilikher,  J. Opt. Soc. of America B \textbf{21}, 113 (2004); K. Yu. Bliokh, Yu. P. Bliokh, V. Freilikher, A. Z. Genack, B. Hu, and P. Sebbah, Phys. Rev. Lett. \textbf{97}, 243904 (2006);  K.Y. Bliokh, Y.P. Bliokh, V. Freilikher, A.Z. Genack, and P. Sebbah,  Phys. Rev. Lett. \textbf{101}, 133901 (2008);  I.V. Shadrivov, K.Y. Bliokh, Y.S. Kivshar, Y.P. Bliokh, and V. Freilikher,  Phys. Rev. Lett. \textbf{104}, 123902 (2010); K.Y. Bliokh, S.A. Gredeskul, P. Rajan, I.V. Shadrivov, and Y.S. Kivshar, Phys. Rev. B \textbf{85}, 014205 (2012).


\bibitem{Edelman} {A. Edelman and E. Kostlan, Bull. Am. Math. Soc. \textbf{32}, No. 1, 1-37 (1995), p.14.}

\bibitem{Avishai} A.Y. Avishai and Y. Band, Phys. Rev. B \textbf{32}, 2674 (1985).

\bibitem{Akkermans} B.E. Akkermans, G. Dunne, and E. Levy, arXiv:1210.7409v2 (2012).

\bibitem{Settimi} A. Settimi, S. Severini, N. Mattiucci, C. Sibilia, M.
Centini, G. D'Aguanno, and M. Bertolotti, Phys. Rev. E \textbf{68}, 026614
(2003).

\bibitem{GR} {I.S. Gradshteyn and I.M. Ryzhik, \textit{Table of Integrals,
Series, and Products}, Seventh Edition, Academic Press, 2007, pp.1,2}
































 











\end{thebibliography}
\end{document}